\documentclass[11pt, letterpaper]{article}
\usepackage[utf8]{inputenc}
\usepackage[margin=1.5cm]{geometry}
\usepackage{titlesec}
\usepackage{tabu}
\usepackage{scalerel}
\usepackage{svg}
\usepackage{enumitem}
\usepackage{amssymb}
\usepackage{xcolor}
\newlist{selectlist}{itemize}{2}
\setlist[selectlist]{label=$\square$,leftmargin=*,noitemsep,topsep=0pt}

\usepackage{lmodern}

\usepackage{hyperref}
\hypersetup{
    colorlinks=true,
    linkcolor=blue,
    filecolor=magenta,      
    urlcolor=blue,
}

\usepackage{amsthm}
\usepackage{stmaryrd}
\usepackage{tikz}
\usepackage[braket, qm]{qcircuit}
\usepackage{graphicx}
\usepackage{csquotes}
\usepackage{listings}
\usepackage[style=ieee, maxnames=6, minnames=1, date=year, doi=false,isbn=false,backend=biber, sortcites, dashed=false]{biblatex}
\AtEveryBibitem{
  \clearname{editor}%
  \clearfield{series}%
  \clearfield{isbn}%
  \clearfield{issn}%
  \clearfield{urldate}%
  \clearfield{url}%
  \clearfield{urldate}%
}

\newlength{\myMheight}

\titleformat{\section}[block]{\hspace{1em}\bfseries}{\thesection.}{0.5em}{} 
\titleformat{\subsection}[block]{\hspace{1em}}{\thesubsection}{0.5em}{}

\usepackage[textsize=footnotesize]{todonotes}
\theoremstyle{definition}

\begin{document}

\noindent
\textbf{\textit{Title / name of your software}}\\\noindent
SyReC Synthesizer: An MQT Tool for Synthesis of Reversible Circuits
\vskip0.5cm
\noindent
\textbf{\textit{Names of authors / main developers (incl. affiliations, addresses, email)}}\\
Smaran Adarsh \emph{(Corresponding Author)}\\
Chair for Design Automation, Technical University of Munich, Arcisstraße 21, 80333 Munich, Germany\\
smaran.adarsh@tum.de
\vskip0.5cm\noindent
Lukas Burgholzer\\
Institute for Integrated Circuits, Johannes Kepler University, Altenberger Straße 69, 4040 Linz, Austria\\
lukas.burgolzer@jku.at
\vskip0.5cm\noindent
Tanmay Manjunath\\
Department of Microelectronics and Computer Engineering, Delft University of Technology, Mekelweg 5, 2628 CD Delft, The Netherlands
\vskip0.5cm\noindent
Robert Wille\\
Chair for Design Automation, Technical University of Munich, Arcisstraße 21, 80333 Munich, Germany\\
Software Competence Center Hagenberg GmbH (SCCH), Softwarepark 32a, 4232 Hagenberg, Austria\\
robert.wille@tum.de
\vskip0.5cm

\noindent
\textbf{Abstract}\\
Reversible circuits form the backbone for many promising emerging technologies such as quantum computing, low power/adiabatic design, encoder/decoder devices, and several other applications. 
In the recent years, the scalable synthesis of such circuits has gained significant attention. 
In this work, we present the \emph{SyReC Synthesizer}, a synthesis tool for reversible circuits based on the hardware description language SyReC. SyReC allows to describe reversible functionality at a high level of abstraction. 
The provided \emph{SyReC Synthesizer} then realizes this functionality in a push-button fashion. Corresponding options allow for a trade-off between the number of needed circuit signals/lines (relevant, e.g., for quantum computing in which every circuit line corresponds to a qubit) and the respectively needed gates (corresponding to the circuit's costs). 
Furthermore, the tool allows to simulate the resulting circuit as well as to determine the gate costs of it.
The \emph{SyReC Synthesizer} is available as an open-source software package at \url{https://github.com/cda-tum/syrec} as part of the Munich Quantum Toolkit (MQT).
\vskip0.5cm

\newpage

\noindent
\textbf{Code metadata}\\

\begin{table}[!h]
\begin{tabular}{|l|p{6.5cm}|p{8.5cm}|}
\hline
\textbf{Nr.} & \textbf{Code metadata description} & \textbf{Please fill in this column} \\
\hline
C1 & Current code version & v1.0.0 \\
\hline
C2 & Permanent link to code/repository used for this code version & \url{https://github.com/cda-tum/syrec} \\
\hline
C3  & Permanent link to Reproducible Capsule & \url{https://codeocean.com/capsule/9148434/tree/v1} \\
\hline
C4 & Legal Code License   & MIT \\
\hline
C5 & Code versioning system used & git \\
\hline
C6 & Software code languages, tools, and services used & C++, Python, Qt \\
\hline
C7 & Compilation requirements, operating environments \& dependencies & cmake $\ge$ 3.14, C++-17 compatible compiler, Python $\ge$ 3.7\\
\hline
C8 & If available Link to developer documentation/manual & \url{https://syrec.readthedocs.io} \\
\hline
C9 & Support email for questions & quantum.cda@xcit.tum.de\\
\hline
\end{tabular}
\caption{Code metadata (mandatory)}
\label{} 
\end{table}
\vskip0.5cm

\section{Introduction}\label{sec:Introduction}

Reversible circuits belong to a class of emerging computing devices that realize invertible Boolean functions, i.e., functions in which each input pattern is mapped to a unique output pattern and vice versa. This reversible characteristic is essential for the development of many emerging technologies such as quantum computing ~\cite{nielsenQuantumComputationQuantum2010}, adiabatic circuits~\cite{zulehnerDesignAutomationAdiabatic2019, berutExperimentalVerificationLandauer2012}, encoding and decoding devices~\cite{zulehnerTakingOnetooneMappings2017, zulehnerExploitingCodingTechniques2018}, verification~\cite{burgholzerCharacteristicsReversibleCircuits2022, burgholzerExploitingReversibleComputing2022}, as well as many others.

Accordingly, the design and synthesis of reversible circuits received significant attention in the past. Originally, corresponding synthesis approaches primarily relied on a functional description provided at a low abstraction level such as truth tables~\cite{millerTransformationBasedAlgorithm2003,willeQuantifiedSynthesisReversible2008}, ESOP-based representations~\cite{fazelESOPbasedToffoliGate2007}, or decision diagrams~\cite{willeBDDbasedSynthesisReversible2009, soekenSynthesisReversibleCircuits2012}. However, to efficiently design and realize complex applications, description means at higher levels of abstractions are crucial. Accordingly, researchers started the development of reversible \emph{Hardware Description Languages}~(HDLs) that allow for describing the intended functionality at a high level of abstraction while, at the same, ensuring the reversibility throughout the design~\cite{willeSyReCProgrammingLanguage2010, thomsenFunctionalLanguageDescribing2012}. 

In this work, we present the \emph{SyReC Synthesizer}, an open-source implementation of the synthesis approach for the reversible HDL SyReC~\cite{willeSyReCProgrammingLanguage2010}. The \emph{SyReC Synthesizer} is a part of the \emph{Munich Quantum Toolkit} (MQT) and provides two different synthesis schemes as proposed in~\cite{willeSyReCProgrammingLanguage2010, willeSyReCHardwareDescription2016, willeHDLbasedSynthesisReversible2019} (each coming with certain advantages and disadvantages). The tool also provides an \mbox{easy-to-use} \emph{Graphical User Interface}~(GUI) 
for users to conveniently synthesize reversible circuits from corresponding SyReC descriptions.

\section{Description and Features}\label{sec:features}

The \emph{SyReC Synthesizer} allows users to automatically synthesize reversible circuits from a high-level HDL description~\cite{willeSyReCHardwareDescription2016}. 
The proposed tool is mainly implemented in C++ and works on all major operating systems. Moreover, with the help of Python bindings, the tool provides an easy-to-use GUI which allows users to specify the desired functionality similar to a classical IDE.
The tool can easily be installed (without any need for compilation) and started by running the following commands:
\begin{lstlisting}[]
	pip install mqt.syrec
	syrec-editor #To start the SyReC editor GUI
\end{lstlisting}

Afterwards, the tool accepts any HDL description following the SyReC  

grammar and syntax as described in detail in~\cite{willeSyReCHardwareDescription2016}.
As an example, consider the SyReC program for a simple 2-bit \emph{Arithmetic Logic Unit}~(ALU) as shown in \autoref{fig:guiALU}. This ALU consists of an input signal \texttt{op} based on which either the sum or the difference of the inputs \texttt{x1} and \texttt{x2} is stored in the output signal \texttt{x0}. By clicking on the  \scalerel*{\includegraphics{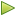}}{1.0} button, the tool automatically synthesizes the corresponding circuit. To this end, two complementary synthesis schemes are available:

\begin{figure}[t]
    \centering
    \includegraphics[width=0.50\textwidth]{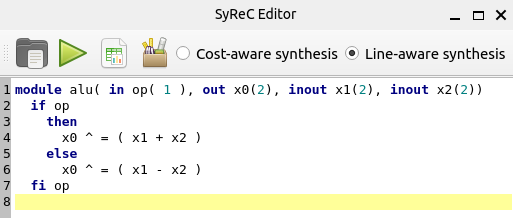}
    \caption{SyReC ALU program along with the provided GUI.}
    \label{fig:guiALU}
\end{figure}

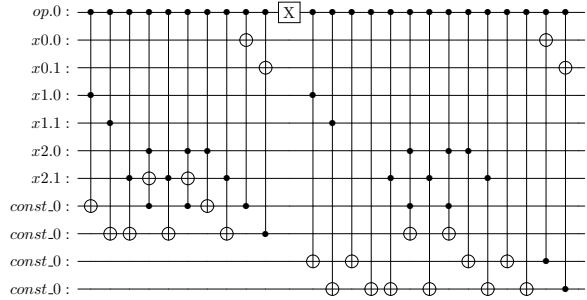
\begin{figure*}[t]
    \centering
\scalebox{0.55}{
\Qcircuit @C=0.4em @R=0.45em @!R { \\
	 	\nghost{{q}_{0} :  } & \lstick{op.0 :  } & \ctrl{3} & \ctrl{4} & \ctrl{6} & \ctrl{5} & \ctrl{6} & \ctrl{5} & \ctrl{5} & \ctrl{6} & \ctrl{1} & \ctrl{2} & \gate{\mathrm{X}} & \ctrl{3} & \ctrl{4} & \ctrl{9} & \ctrl{10} & \ctrl{6} & \ctrl{5} & \ctrl{6} & \ctrl{5} & \ctrl{5} & \ctrl{6} & \ctrl{9} & \ctrl{10} & \ctrl{1} & \ctrl{2} & \qw & \qw\\
	 	\nghost{{q}_{1} :  } & \lstick{x0.0 : } & \qw & \qw & \qw & \qw & \qw & \qw & \qw & \qw & \targ & \qw & \qw & \qw & \qw & \qw & \qw & \qw & \qw & \qw & \qw & \qw & \qw & \qw & \qw & \targ & \qw & \qw & \qw\\
	 	\nghost{{q}_{2} :  } & \lstick{x0.1 : } & \qw & \qw & \qw & \qw & \qw & \qw & \qw & \qw & \qw & \targ & \qw & \qw & \qw & \qw & \qw & \qw & \qw & \qw & \qw & \qw & \qw & \qw & \qw & \qw & \targ & \qw & \qw\\
	 	\nghost{{q}_{3} :  } & \lstick{x1.0 :  } & \ctrl{4} & \qw & \qw & \qw & \qw & \qw & \qw & \qw & \qw & \qw & \qw & \ctrl{6} & \qw & \qw & \qw & \qw & \qw & \qw & \qw & \qw & \qw & \qw & \qw & \qw & \qw & \qw & \qw\\
	 	\nghost{{q}_{4} :  } & \lstick{x1.1 :  } & \qw & \ctrl{4} & \qw & \qw & \qw & \qw & \qw & \qw & \qw & \qw & \qw & \qw & \ctrl{6} & \qw & \qw & \qw & \qw & \qw & \qw & \qw & \qw & \qw & \qw & \qw & \qw & \qw & \qw\\
	 	\nghost{{q}_{5} :  } & \lstick{x2.0 :  } & \qw & \qw & \qw & \ctrl{1} & \qw & \ctrl{1} & \ctrl{2} & \qw & \qw & \qw & \qw & \qw & \qw & \qw & \qw & \qw & \ctrl{2} & \qw & \ctrl{2} & \ctrl{4} & \qw & \qw & \qw & \qw & \qw & \qw & \qw\\
	 	\nghost{{q}_{6} :  } & \lstick{x2.1 :  } & \qw & \qw & \ctrl{2} & \targ & \ctrl{2} & \targ & \qw & \ctrl{2} & \qw & \qw & \qw & \qw & \qw & \qw & \qw & \ctrl{4} & \qw & \ctrl{4} & \qw & \qw & \ctrl{4} & \qw & \qw & \qw & \qw & \qw & \qw\\
	 	\nghost{{q}_{7} :  } & \lstick{{const}\_{0}:  } & \targ & \qw & \qw & \ctrl{-1} & \qw & \ctrl{-1} & \targ & \qw & \ctrl{-6} & \qw & \qw & \qw & \qw & \qw & \qw & \qw & \ctrl{1} & \qw & \ctrl{1} & \qw & \qw & \qw & \qw & \qw & \qw & \qw & \qw\\
	 	\nghost{{q}_{8} :  } & \lstick{{const}\_{0}:  } & \qw & \targ & \targ & \qw & \targ & \qw & \qw & \targ & \qw & \ctrl{-6} & \qw & \qw & \qw & \qw & \qw & \qw & \targ & \qw & \targ & \qw & \qw & \qw & \qw & \qw & \qw & \qw & \qw\\
	 	\nghost{{q}_{9} :  } & \lstick{{const}\_{0}:  } & \qw & \qw & \qw & \qw & \qw & \qw & \qw & \qw & \qw & \qw & \qw & \targ & \qw & \targ & \qw & \qw & \qw & \qw & \qw & \targ & \qw & \targ & \qw & \ctrl{-8} & \qw & \qw & \qw\\
	 	\nghost{{q}_{10} :  } & \lstick{{const}\_{0}:  } & \qw & \qw & \qw & \qw & \qw & \qw & \qw & \qw & \qw & \qw & \qw & \qw & \targ & \qw & \targ & \targ & \qw & \targ & \qw & \qw & \targ & \qw & \targ & \qw & \ctrl{-8} & \qw & \qw\\
\\ }}
\caption{ALU circuit resulting from cost-aware synthesis.}
	 \label{fig:aluWithAdditionalLines}
    \end{figure*}

\begin{figure*}[t]
    \centering
\scalebox{0.55}{\Qcircuit @C=0.3em @R=0.75em @!R { \\
	 	\nghost{{q}_{0} :  } & \lstick{op.0 :  } & \ctrl{4} & \ctrl{3} & \ctrl{4} & \ctrl{3} & \ctrl{3} & \ctrl{4} & \ctrl{1} & \ctrl{2} & \ctrl{5} & \ctrl{6} & \ctrl{4} & \ctrl{3} & \ctrl{4} & \ctrl{3} & \ctrl{3} & \ctrl{4} & \ctrl{5} & \ctrl{6} & \gate{\mathrm{X}} & \ctrl{3} & \ctrl{4} & \ctrl{4} & \ctrl{3} & \ctrl{4} & \ctrl{3} & \ctrl{3} & \ctrl{4} & \ctrl{3} & \ctrl{4} & \ctrl{5} & \ctrl{6} & \ctrl{1} & \ctrl{2} & \ctrl{3} & \ctrl{4} & \ctrl{4} & \ctrl{3} & \ctrl{4} & \ctrl{3} & \ctrl{3} & \ctrl{4} & \ctrl{3} & \ctrl{4} & \ctrl{5} & \ctrl{6} & \qw\\
	 	\nghost{{q}_{1} :  } & \lstick{x0.0 :  } & \qw & \qw & \qw & \qw & \qw & \qw & \targ & \qw & \qw & \qw & \qw & \qw & \qw & \qw & \qw & \qw & \qw & \qw & \qw & \qw & \qw & \qw & \qw & \qw & \qw & \qw & \qw & \qw & \qw & \qw & \qw & \targ & \qw & \qw & \qw & \qw & \qw & \qw & \qw & \qw & \qw & \qw & \qw & \qw & \qw & \qw\\
	 	\nghost{{q}_{2} :  } & \lstick{x0.1 :  } & \qw & \qw & \qw & \qw & \qw & \qw & \qw & \targ & \qw & \qw & \qw & \qw & \qw & \qw & \qw & \qw & \qw & \qw & \qw & \qw & \qw & \qw & \qw & \qw & \qw & \qw & \qw & \qw & \qw & \qw & \qw & \qw & \targ & \qw & \qw & \qw & \qw & \qw & \qw & \qw & \qw & \qw & \qw & \qw & \qw & \qw\\
	 	\nghost{{q}_{3} :  } & \lstick{x1.0 :  } & \qw & \ctrl{1} & \qw & \ctrl{1} & \ctrl{2} & \qw & \qw & \qw & \qw & \qw & \qw & \ctrl{1} & \qw & \ctrl{1} & \ctrl{2} & \qw & \qw & \qw & \qw & \targ & \qw & \qw & \ctrl{1} & \qw & \ctrl{1} & \ctrl{2} & \qw & \targ & \qw & \qw & \qw & \qw & \qw & \targ & \qw & \qw & \ctrl{1} & \qw & \ctrl{1} & \ctrl{2} & \qw & \targ & \qw & \qw & \qw & \qw\\
	 	\nghost{{q}_{4} :  } & \lstick{x1.1 :  } & \ctrl{2} & \targ & \ctrl{2} & \targ & \qw & \ctrl{2} & \qw & \qw & \qw & \qw & \ctrl{2} & \targ & \ctrl{2} & \targ & \qw & \ctrl{2} & \qw & \qw & \qw & \qw & \targ & \ctrl{2} & \targ & \ctrl{2} & \targ & \qw & \ctrl{2} & \qw & \targ & \qw & \qw & \qw & \qw & \qw & \targ & \ctrl{2} & \targ & \ctrl{2} & \targ & \qw & \ctrl{2} & \qw & \targ & \qw & \qw & \qw\\
	 	\nghost{{q}_{5} :  } & \lstick{x2.0 :  } & \qw & \ctrl{-1} & \qw & \ctrl{-1} & \targ & \qw & \ctrl{-4} & \qw & \targ & \qw & \qw & \ctrl{-1} & \qw & \ctrl{-1} & \targ & \qw & \targ & \qw & \qw & \qw & \qw & \qw & \ctrl{-1} & \qw & \ctrl{-1} & \targ & \qw & \qw & \qw & \targ & \qw & \ctrl{-4} & \qw & \qw & \qw & \qw & \ctrl{-1} & \qw & \ctrl{-1} & \targ & \qw & \qw & \qw & \targ & \qw & \qw\\
	 	\nghost{{q}_{6} :  } & \lstick{x2.1 :  } & \targ & \qw & \targ & \qw & \qw & \targ & \qw & \ctrl{-4} & \qw & \targ & \targ & \qw & \targ & \qw & \qw & \targ & \qw & \targ & \qw & \qw & \qw & \targ & \qw & \targ & \qw & \qw & \targ & \qw & \qw & \qw & \targ & \qw & \ctrl{-4} & \qw & \qw & \targ & \qw & \targ & \qw & \qw & \targ & \qw & \qw & \qw & \targ & \qw\\
\\ }}
\caption{ALU circuit resulting from line-aware synthesis}
	 \label{fig:aluWithoutAdditionalLines}
    \end{figure*}
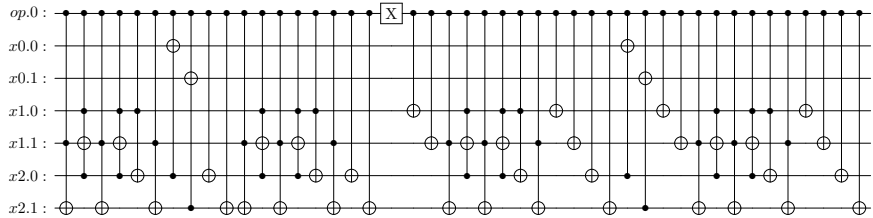

\begin{enumerate}
	\item \textit{Cost-aware synthesis:} \label{scheme:costAware}
	
	In this synthesis scheme, additional circuit lines (representing circuit signals) are introduced to store the corresponding intermediate results of each operation---thereby providing the liberty to accordingly modify the circuit without having any effect on the original inputs. This results in a circuit description where
	the gate costs\footnote{Gate cost is defined as the number of elementary reversible operations required to realize a gate.} are kept moderate at the expense of a substantially larger number of additionally needed circuit lines.
	
	\autoref{fig:aluWithAdditionalLines} depicts the circuit resulting for
	the SyReC ALU description from \autoref{fig:guiALU} using this cost-aware synthesis scheme.
	
	\item \textit{Line-aware synthesis:}
	
	In this synthesis scheme, 
	the goal is to keep the number of additionally required circuit lines as small as possible. To this end, the corresponding intermediate results of each operation are computed and stored using one of the available circuit lines (rather than additional circuit lines). Afterwards, the inputs are re-computed with the help of the corresponding inverse operations (see~\cite{willeSyReCHardwareDescription2016} for details). As this substantially increases the number of required gates and, hence, the gate costs of the circuit, dedicated building blocks and  
	re-writing as well as translation techniques~\cite{willeCircuitLineMinimization2012, al-wardiRewritingHDLDescriptions2016, willeHDLbasedSynthesisReversible2019} are additionally employed to keep this overhead in check.  
	\autoref{fig:aluWithoutAdditionalLines} depicts the circuit resulting for
	the SyReC ALU description from \autoref{fig:guiALU} using this line-aware synthesis scheme.
\end{enumerate}

\autoref{fig:aluWithAdditionalLines} and \autoref{fig:aluWithoutAdditionalLines} clearly show the differences in the gate costs and the number of lines of the resulting circuits: While the line-aware synthesis scheme generates a circuit with no additional circuit lines, it is quite inefficient concerning the number of gates. Vice versa, the circuit generated by the cost-aware synthesis scheme keeps the number of gates moderate, but results in a substantial amount of additional circuit lines. 
Depending on the addressed technology, these differences can have a huge impact. 
In quantum computing, for example, the number of circuit lines corresponds to the number of qubits---a rather restricted resource.
Thus, in this case, it is desirable to keep the number of additional circuit lines as small as possible. If this is not the case, the gate costs are obviously the primary objective and a larger number of circuit lines can easily be accepted. 
By providing both synthesis schemes, the \emph{SyReC Synthesizer} allows users to choose the method that best fits the addressed technology.

In addition to that, the tool allows users to quickly evaluate the costs of the resulting circuits (by clicking the \scalerel*{\includegraphics{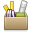}}{1.0} button) and offers a simulation method to check the resulting circuits' functionality (by clicking the \scalerel*{\includegraphics{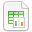}}{1.0} button).

\section{Impact Overview}\label{sec:Impact}

Reversible circuits provide a main basis for many emerging technologies and promising applications. Prominent examples include, e.g.,
\begin{itemize}
    \item quantum computing~\cite{nielsenQuantumComputationQuantum2010} in which building blocks and so-called oracles rely on reversible circuits,
    \item adiabatic circuits that utilizes reversible computations as demonstrated, e.g., in~\cite{zulehnerDesignAutomationAdiabatic2019},
    \item encoders and decoders~\cite{zulehnerTakingOnetooneMappings2017, zulehnerExploitingCodingTechniques2018} which can benefit from reversible circuits as they naturally realize one-to-one mappings,
    \item verification, in which characteristics of reversible logic can be exploited as proposed in~\cite{burgholzerCharacteristicsReversibleCircuits2022, burgholzerExploitingReversibleComputing2022}, and
    \item many more.
\end{itemize}

All these application areas would substantially benefit from reversible circuits that can be described and, afterwards, are realized in terms of \emph{Hardware Description Languages} (HDLs)---lifting the design process to higher levels of abstraction. However, while research on HDL-based synthesis of reversible circuits has been conducted for many years, barely any software tools and methods are publicly available. Not only does this slow down the research progress (since it becomes increasingly hard to compare against existing methods), it also means that barely anyone can explore this new and vastly different paradigm of developing circuits and systems.

The \emph{SyReC Synthesizer} has been developed to change this dire situation.
To this end, it provides open-source implementations of several key components in the HDL-based synthesis flow:
\begin{itemize}
	\item A reference implementation of the SyReC grammar together with a corresponding parser.
	\item Implementations of the state-of-the-art \mbox{cost-aware} and \mbox{line-aware} synthesis methods described above.
	\item An \mbox{easy-to-use} GUI for specifying, synthesizing, simulating, and analyzing SyReC programs and the resulting circuit realizations.
\end{itemize}
As a result, the proposed tool lays the foundation for developing and incorporating even more methods for the design of reversible circuits and systems in a \mbox{publicly-available} software package.
Furthermore, by exposing all functionality via an easy to use Python interface and distributing the tool as a (pre-compiled) Python package, reversible computing becomes way more accessible to a much broader audience.

\section{Conclusions}\label{sec:conclusions}
In this paper, we presented the HDL-based reversible circuit synthesis tool \emph{SyReC Synthesizer} which is part of the \emph{Munich Quantum Toolkit} (MQT). 
The tool offers an easy-to-use GUI which allows users to conveniently specify SyReC descriptions and, afterwards, to automatically synthesize corresponding reversible circuits out of it. 
To this end, two different schemes focusing on gate costs and number of lines, respectively, are provided. The tool is available as an open-source software package at \url{https://github.com/cda-tum/syrec}.
An elaborate description of the different synthesis schemes is available in~\cite{willeSyReCProgrammingLanguage2010, willeSyReCHardwareDescription2016, willeHDLbasedSynthesisReversible2019}. 

\vskip0.3cm\noindent
\textbf{Acknowledgements}\\
This work received funding from the European Research Council (ERC) under the European Union’s Horizon 2020 research and innovation program (grant agreement No. 101001318), was part of the Munich Quantum Valley, which is supported by the Bavarian state government with funds from the Hightech Agenda Bayern Plus, and has been supported by the BMWK on the basis of a decision by the German Bundestag through project QuaST, as well as by the BMK, BMDW, and the State of Upper Austria in the frame of the COMET program (managed by the FFG).

\printbibliography

@inproceedings{al-wardiRewritingHDLDescriptions2016,
  title = {Re-writing {{HDL}} descriptions for line-aware synthesis of reversible circuits},
  booktitle = ismvl,
  author = {Al-Wardi, Zaid and Wille, Robert and Drechsler, Rolf},
  date = {2016},
  pages = {31--36},
  doi = {10.1109/ISMVL.2016.36},
  eventtitle = {ismvl}
}

@article{berutExperimentalVerificationLandauer2012,
  title = {Experimental verification of {{Landauer}}’s principle linking information and thermodynamics},
  author = {Bérut, Antoine and Arakelyan, Artak and Petrosyan, Artyom and Ciliberto, Sergio and Dillenschneider, Raoul and Lutz, Eric},
  date = {2012-03},
  journaltitle = {Nature},
  volume = {483},
  number = {7388},
  pages = {187--189}
}

@article{burgholzerCharacteristicsReversibleCircuits2022,
  title = {Characteristics of reversible circuits for error detection},
  author = {Burgholzer, Lukas and Wille, Robert and Kueng, Richard},
  date = {2022-04},
  journaltitle = {Array},
  eprint = {2012.02037},
  eprinttype = {arxiv},
  primaryclass = {cs},
  archiveprefix = {arXiv}
}

@inproceedings{burgholzerExploitingReversibleComputing2022,
  title = {Exploiting reversible computing for verification: {{Potential}}, possible paths, and consequences},
  booktitle = aspdac,
  author = {Burgholzer, Lukas and Wille, Robert},
  date = {2022},
  eventtitle = {aspdac}
}

@inproceedings{fazelESOPbasedToffoliGate2007,
  title = {{{ESOP-based Toffoli}} gate cascade generation},
  booktitle = pacrim,
  author = {Fazel, Kenneth and Thornton, Mitchell A and Rice, Jacqueline E},
  date = {2007},
  pages = {206--209},
  eventtitle = {pacrim}
}

@inproceedings{millerTransformationBasedAlgorithm2003,
  title = {A transformation based algorithm for reversible logic synthesis},
  booktitle = dac,
  author = {Miller, D. Michael and Maslov, Dmitri and Dueck, Gerhard W.},
  date = {2003},
  pages = {318--323},
  doi = {10.1145/775832.775915},
  eventtitle = {dac}
}

@book{nielsenQuantumComputationQuantum2010,
  title = {Quantum {{Computation}} and {{Quantum Information}}},
  shorttitle = {Quantum {{Computation}} and {{Quantum Information}}},
  author = {Nielsen, Michael A. and Chuang, Isaac L.},
  date = {2010},
  publisher = {{Cambridge University Press}}
}

@inproceedings{soekenSynthesisReversibleCircuits2012,
  title = {Synthesis of reversible circuits with minimal lines for large functions},
  booktitle = aspdac,
  author = {Soeken, M. and Wille, R. and Hilken, C. and Przigoda, N. and Drechsler, R.},
  date = {2012},
  pages = {85--92},
  eventtitle = {aspdac}
}

@inproceedings{thomsenFunctionalLanguageDescribing2012,
  title = {A functional language for describing reversible logic},
  booktitle = fdl,
  author = {Thomsen, Michael Kirkedal},
  date = {2012},
  pages = {135--142},
  eventtitle = {fdl}
}

@inproceedings{willeBDDbasedSynthesisReversible2009,
  title = {{{BDD-based}} synthesis of reversible logic for large functions},
  booktitle = dac,
  author = {Wille, Robert and Drechsler, Rolf},
  date = {2009},
  pages = {270--275},
  doi = {10.1145/1629911.1629984},
  eventtitle = {dac}
}

@inproceedings{willeCircuitLineMinimization2012,
  title = {Circuit line minimization in the {{HDL-based}} synthesis of reversible logic},
  booktitle = isvlsi,
  author = {Wille, Robert and Soeken, Mathias and Schönborn, Eleonora and Drechsler, Rolf},
  date = {2012},
  pages = {213--218},
  doi = {10.1109/ISVLSI.2012.43},
  eventtitle = {isvlsi}
}

@inproceedings{willeHDLbasedSynthesisReversible2019,
  title = {Towards {{HDL-based}} synthesis of reversible circuits with no additional lines},
  booktitle = iccad,
  author = {Wille, Robert and Haghparast, Majid and Adarsh, Smaran and Tanmay, M.},
  date = {2019},
  doi = {10.1109/ICCAD45719.2019.8942156},
  eventtitle = {iccad}
}

@inproceedings{willeQuantifiedSynthesisReversible2008,
  title = {Quantified synthesis of reversible logic},
  booktitle = date,
  author = {Wille, Robert and Le, Hoang Minh and Dueck, Gerhard W. and Große, Daniel},
  date = {2008},
  pages = {1015--1020},
  doi = {10.1109/DATE.2008.4484814},
  eventtitle = {date}
}

@article{willeSyReCHardwareDescription2016,
  title = {{{SyReC}}: {{A}} hardware description language for the specification and synthesis of reversible circuits},
  shorttitle = {{{SyReC}}},
  author = {Wille, Robert and Schönborn, Eleonora and Soeken, Mathias and Drechsler, Rolf},
  date = {2016-03},
  journaltitle = {Integration},
  volume = {53},
  pages = {39--53}
}

@inproceedings{willeSyReCProgrammingLanguage2010,
  title = {{{SyReC}}: {{A}} programming language for synthesis of reversible circuits},
  booktitle = fdl,
  author = {Wille, R. and Offermann, S. and Drechsler, R.},
  date = {2010},
  eventtitle = {fdl}
}

@inproceedings{zulehnerDesignAutomationAdiabatic2019,
  title = {Design automation for adiabatic circuits},
  booktitle = aspdac,
  author = {Zulehner, Alwin and Frank, Michael P. and Wille, Robert},
  date = {2019},
  pages = {669--674},
  publisher = {{ACM}},
  doi = {10.1145/3287624.3287673},
  eventtitle = {aspdac}
}

@inproceedings{zulehnerExploitingCodingTechniques2018,
  title = {Exploiting coding techniques for logic synthesis of reversible circuits},
  booktitle = aspdac,
  author = {Zulehner, Alwin and Wille, Robert},
  date = {2018},
  pages = {670--675},
  doi = {10.1109/ASPDAC.2018.8297399},
  eventtitle = {aspdac}
}

@inproceedings{zulehnerTakingOnetooneMappings2017,
  title = {Taking one-to-one mappings for granted: {{Advanced}} logic design of encoder circuits},
  shorttitle = {Taking one-to-one mappings for granted},
  booktitle = date,
  author = {Zulehner, Alwin and Wille, Robert},
  date = {2017},
  pages = {818--823},
  doi = {10.23919/DATE.2017.7927101},
  eventtitle = {date}
}

@STRING{iccad	= {Int'l Conf. on CAD} }

@STRING{isvlsi	= {IEEE Annual Symp. on VLSI} }

@STRING{dac	= {Design Automation Conf.} }

@STRING{aspdac	= {Asia and South Pacific Design Automation Conf.} }

@STRING{date	= {Design, Automation and Test in Europe} }

@STRING{fdl	= {Forum on Specification and Design Languages} }

@STRING{ismvl	= {Int'l Symp. on {M}ulti-{V}alued {L}ogic} }

@STRING{pacrim  = {Pacific Rim Conference on Communications, Computers and 
		  Signal Processing} }

\end{document}